\documentclass[prd,twocolumn,reprint,superscriptaddress,showpac]{revtex4}
\usepackage{feynmp,amsmath,amssymb,graphicx,subfigure,hyperref,bm,mathrsfs,slashed}
\usepackage{CJK}
\usepackage{color}

\begin{document}
\begin{CJK*}{GBK}{song}

\title{Anisotropic longitudinal optical conductivities of tilted Dirac bands in 1T$^\prime$-MoS$_2$}

\author{Chao-Yang Tan}
\thanks{These authors have contributed equally to this work.}
\affiliation{Department of Physics, Institute of Solid State Physics and
Center for Computational Sciences, Sichuan Normal University, Chengdu,
Sichuan 610066, China}

\author{Chang-Xu Yan}
\thanks{These authors have contributed equally to this work.}
\affiliation{Department of Physics, Institute of Solid State Physics and
Center for Computational Sciences, Sichuan Normal University, Chengdu,
Sichuan 610066, China}

\author{Yong-Hong Zhao}
\affiliation{Department of Physics, Institute of Solid State Physics and
Center for Computational Sciences, Sichuan Normal University, Chengdu,
Sichuan 610066, China}

\author{Hong Guo}
\affiliation{Department of Physics, Institute of Solid State Physics and
Center for Computational Sciences, Sichuan Normal University, Chengdu,
Sichuan 610066, China}
\affiliation{Department of Physics, McGill University, Montreal, Quebec H3A 2T8, Canada}

\author{Hao-Ran Chang}
\thanks{Corresponding author:hrchang@mail.ustc.edu.cn}
\affiliation{Department of Physics, Institute of Solid State Physics and
Center for Computational Sciences, Sichuan Normal University, Chengdu,
Sichuan 610066, China}
\affiliation{Department of Physics, McGill University, Montreal, Quebec H3A 2T8, Canada}

\date{\today}

\begin{abstract}
1T$^\prime$-MoS$_2$ exhibits valley-spin-polarized tilted Dirac bands in the presence of external vertical electric field and undergoes a topological phase transition between the topological insulator and band insulator around the critical value of the electric field. Within the linear response theory, we theoretically investigate the anisotropic longitudinal optical conductivities of tilted Dirac bands in both undoped and doped 1T$^\prime$-MoS$_2$, including the effects of the vertical electric field. The influence of the spin-orbit coupling gap, band tilting, and vertical electric field on the optical conductivities of tilted Dirac bands is revealed. A theoretical scheme for probing the topological phase transition in 1T$^\prime$-MoS$_2$ via exotic behaviors of longitudinal optical conductivities is proposed. The results for 1T$^\prime$-MoS$_2$ are expected to be qualitatively valid for other monolayer tilted gapped Dirac materials, such as $\alpha$-SnS$_2$, TaCoTe$_2$, and TaIrTe$_4$, due to the similarity in their band structures.
\end{abstract}

\maketitle

\end{CJK*}

\section{Introduction\label{Sec:intro}}

The discovery of graphene has led to extremely active research in two-dimensional (2D) Dirac materials \cite{Science2004,RMP2009}-materials characterized by Dirac points and linear and/or hyperbolic energy
dispersions in the momentum space, in clear contrast to traditional metals and semiconductors \cite{Mahan}.
So far, many 2D Dirac materials have been experimentally synthesized and theoretically studied, including $\alpha$-(BEDT-TTF)$_2$I$_3$ \cite{JPSJ2006}, silicene \cite{PRBSilicene2007,PRLSilicene2009,PRBSilicene2011,
PRLSilicene2011,EzawaPRLSilicene2012}, 8-$Pmmn$ borophene \cite{Science8Pmmn2015,PRB8PmmnRapid2016,PRB8Pmmn2016},
and various monolayer transition metal dichalcogenides (TMDCs) \cite{PRLMoS2010,PRLMoS2012,Science2014,AdvMater2016,
NatPhys2017,NatCommun2017,NatPhysTang2017,ScienceWu2018,ScienceFatemi2018,ScienceSajadi2018}. In Dirac materials,
the Dirac dispersion together with the band gap and/or band tilting can be exploited to manipulate physical properties. For example, the valley-spin-polarized band gap tuned by a vertical electric field in silicene generates a topological phase transition between the topological insulator phase and the band insulator phase \cite{EzawaPRLSilicene2012};
the band tilting in 8-$Pmmn$ borophene induces a strong anisotropy in physical properties \cite{PRB8PmmnRapid2016,
PRB8Pmmn2016}. These qualitatively influence important material physics including plasmons \cite{PRBHRC2014,
PRBTabert2014,PRBIurov2017,PRBAgarwal2017,PRBJafari2018,JPSJNishine2011,JPSJNishine2010}, optical conductivities \cite{JPSJNishine2010,PRBStille2012,PRBVerma2017,PRBIurov2018,PRBHerrera2019,PRBIurov2020}, thermoelectric effects \cite{PRBGhosh2020}, Kondo effects \cite{PRBSun2018}, and RKKY interactions \cite{JPCMXiao2014,SciRepWang2018,PRBPaul2019,JMMMZhang2019}.

1T$^\prime$ TMDCs, the monolayer TMDCs of the $\mathrm{T}^\prime$ structure phase, were theoretically predicted to
be quantum spin Hall insulators \cite{Science2014} and have been experimentally synthesized \cite{AdvMater2016,
NatPhys2017,NatCommun2017,NatPhysTang2017,ScienceWu2018,ScienceFatemi2018,ScienceSajadi2018}. Excitingly, the
quantum spin Hall effect \cite{AdvMater2016,NatPhys2017,NatCommun2017,NatPhysTang2017,ScienceWu2018} and gate-induced superconductivity \cite{ScienceFatemi2018,ScienceSajadi2018} were recently observed in experiments. Interestingly, 1T$^\prime$ TMDCs \cite{Science2014} possess tilted Dirac bands around Dirac points similar to that in 8-$Pmmn$ borophene \cite{PRB8PmmnRapid2016,PRB8Pmmn2016}, and undergo a topological phase transition \cite{APLLi2009,PRLMiao2012,PRLZhang2013} induced by a vertical electric field such as that in silicene \cite{EzawaPRLSilicene2012}. The tilted gapped Dirac dispersions \cite{Science2014} are shared by other monolayer Dirac materials, such as $\alpha$-SnS$_2$ \cite{NPGMa2016}, TaCoTe$_2$ \cite{PRBYang2019}, and TaIrTe$_4$ \cite{PRBLu2020}. Furthermore, the band structure of 1T$^\prime$ TMDCs \cite{Science2014} differs strongly from that of its polytypic structure the $1\mathrm{H}$ TMDCs \cite{PRLMoS2010,PRLMoS2012}. These features make the 1T$^\prime$ TMDC a very attractive monolayer Dirac material and its physical properties are worthy of careful investigations.

A particularly important physical property of any material is the optical conductivity whose real part relates to absorption of photons and provides a powerful approach for extracting band structures and optical properties. The optical conductivity of Dirac materials has received intensive research both theoretically and experimentally, for graphene \cite{PRBGusynin2007,PRBStauber2008,PRLMak2008}, $\alpha$-(BEDT-TTF)$_2$I$_3$ \cite{JPSJNishine2010},
silicene \cite{PRBStille2012}, 8-$Pmmn$ borophene \cite{PRBVerma2017}, 1H-MoS$_2$ \cite{PRBCarbotte2012,
PRBCarvalho2013,PRBAsgari2014}, and topological insulators \cite{PRBDiPietro2012,PRBCarbotte2013,PRBXiao2013}.

It is the purpose of this work to investigate the optical conductivity of the monolayer Dirac material 1T$^\prime$
-MoS$_2$ which is a typical material of 1T$^\prime$ TMDCs. This study, apart from further enriching the important materials physics of Dirac materials, is focused on providing a fundamental understanding of at least three outstanding issues. The first is to understand the influence of the spin-orbit coupling (SOC) gap on optical conductivities of tilted Dirac bands, by comparing 1T$^\prime$-MoS$_2$ (gapped) and 8-$Pmmn$ borophene (gapless). The second is to establish
the difference of optical conductivities originating from polytypic structures of MoS$_2$, by comparing 1T$^\prime$-MoS$_2$ with 1H-MoS$_2$; or originating from band tilting, by comparing 1T$^\prime$-MoS$_2$ (tilted) with silicene (untilted). The third is to analyze the probing of the topological phase transition in 1T$^\prime$-MoS$_2$ via longitudinal optical conductivities. To this end, we theoretically investigate the anisotropic longitudinal optical conductivities of tilted Dirac bands in both undoped and doped 1T$^\prime$-MoS$_2$ and the influence of a vertical electric field acting on them.

\begin{figure*}[htbp]
\centering
\includegraphics[width=16cm]{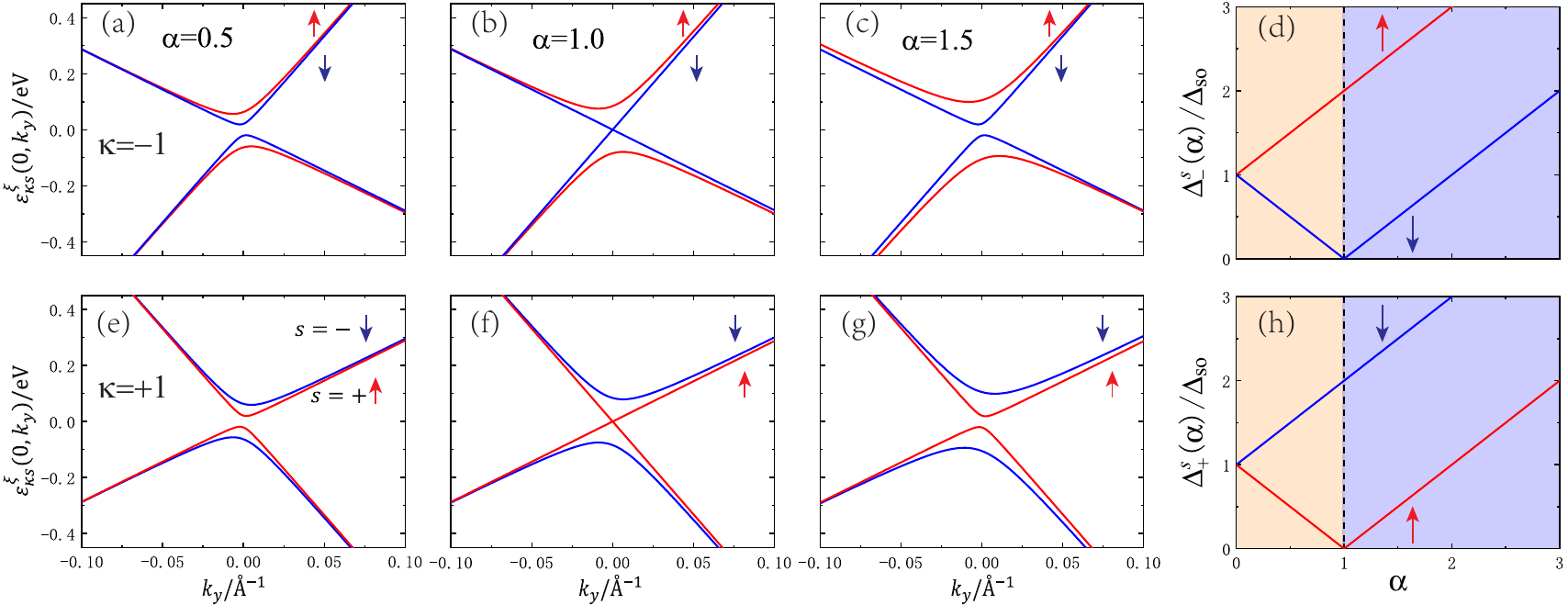}
\caption{(Color online) In the presence of SOC gap and vertical electric field, parametrized respectively by $\Delta_{\mathrm{so}}$ and $\alpha=E/E_c$, the energy bands and gaps are valley-spin-polarized. The upper panels and lower panels stand respectively for $\kappa=-$ valley and $\kappa=+$ valley. The red lines and blue lines denote spin-up ($s=+$) and spin-down ($s=-$), respectively. The topological phase transition between the topological insulator phase ($\alpha<1$) and the band insulator phase ($\alpha>1$) occurs at the phase termed by the valley-spin-polarized metal ($\alpha=1$).}\label{fig1}
\end{figure*}

This paper is organized as follows. In Sec. \ref{Sec:Model}, we briefly describe the model Hamiltonian and formalism
to calculate the longitudinal optical conductivity. In Sec. \ref{Sec:NumericalResults}, we present our numerical results for the absorptive part of the longitudinal conductivity in the undoped and doped 1T$^\prime$-MoS$_2$ when
the vertical electric field is absent and present, respectively. In addition, we provide a detailed analysis of the essential physics. Our main results and conclusions are summarized in Sec. \ref{Sec:Conclusions}. Finally, we give
three appendices to provide key steps of analytical calculation.

\section{Model and theoretical formalism\label{Sec:Model}}

We begin with the low-energy $k\cdot p$ Hamiltonian for 1T$^\prime$-MoS$_2$, reported in Ref. \cite{Science2014}. For simplicity, we further expand it in the vicinity of two independent Dirac points located at $(0,\kappa\Lambda)$ with $\kappa=\pm$. The linearized Hamiltonian around the $\kappa$ valley reads
\begin{align}
H_{\kappa}(k_x,k_y)=&\hbar k_x v_1\gamma_1
+\hbar k_y\left(v_2\gamma_0-\kappa v_{-}I
-\kappa v_{+}\gamma_2\right)
\nonumber\\&
+\Delta_{\mathrm{so}}\left(\kappa\gamma_0-i\alpha\gamma_1\gamma_2\right),
\end{align}
where the wave vector $\boldsymbol{k}=(k_x,k_y)$, the SOC gap $\Delta_{\mathrm{so}}=0.042\mathrm{eV}$, and
the Fermi velocities $v_1=3.87\times10^{5}m/s$, $v_2=0.46\times10^{5}m/s$, $v_{-}=2.86\times10^{5}m/s$, and $v_{+}=7.21\times10^{5}m/s$. In addition, the $4\times4$ unit matrix is given by $I=\tau_0\otimes\sigma_0$; the $4\times4$ Dirac matrices are defined as $\gamma_0=\tau_1\otimes\sigma_1$, $\gamma_1=\tau_2\otimes\sigma_0$, and $\gamma_2=\tau_3\otimes\sigma_0$, in which $\tau_0$ and $\tau_i$ stand for pseudospin space which indicates the conduction and valence band degrees of freedom while $\sigma_0$ and $\sigma_i$ denote Pauli matrices acting upon
real-spin space. The ratio $\alpha$ is defined as $\alpha=|E_{z}/E_{c}|$, where $E_{z}$ is the vertical electric
field and $E_{c}$ denotes its critical value. Hereafter, we set $\hbar=1$ for simplicity.

A straightforward algebra leads to the eigenvalues of this Hamiltonian as
\begin{align}
\varepsilon_{\kappa s}^{\xi}(k_x,k_y)
&=-\kappa v_{-}k_y+ \xi\mathcal{Z}_{\kappa s}(k_x,k_y),
\end{align}
where
\begin{align}
\mathcal{Z}_{\kappa s}(k_x,k_y)
&=\sqrt{\left[A_{\kappa s}(k_y)\right]^2+\left[B(k_x)\right]^2+\left[C_{\kappa}(k_y)\right]^2},
\end{align}
with
\begin{align}
A_{\kappa s}(k_y)&=v_2k_y+(\kappa-\alpha s)\Delta_{\mathrm{so}},\\
B(k_x)&=v_{1}k_x,\\
C_{\kappa}(k_y)&=-\kappa v_{+}k_y.
\end{align}
The indices $s=\pm$ and $\xi=\pm$ denote the spins (up and down) and bands (conduction and valence), respectively.
It is noted that the dispersions are tilted hyperboloids and asymptotically tilted cones in the region of large
wave vector.

In the absence of vertical electric field ($\alpha=0$), the bands are spin-degenerate. However, in the presence of vertical electric field ($\alpha>0$), the spin-up band and spin-down band split at one valley, but reverse their spins at the other valley, which leads to the valley-spin-polarized bands as well as the valley-spin-polarized gaps.
For convenience, we define the valley-spin-polarized gaps as
\begin{align}
\Delta_{\kappa}^{s}(\alpha)=\mathcal{Z}_{\kappa s}(0,0)=|\kappa-\alpha s|\Delta_{\mathrm{so}}.
\end{align}
From these definitions, it is easy to find that the energy bands $\varepsilon_{\kappa s}^{\xi}(k_x,k_y)$ and the
gaps $\Delta_{\kappa}^{s}(\alpha)$ do not change if we substitute $(k_y,\kappa,s)$ with $(-k_y,-\kappa,-s)$. In addition, the dependencies of valley-spin-polarized bands and valley-spin-polarized gaps on the vertical electric
field are explicitly shown in Fig.(\ref{fig1}). Interestingly, at the $\kappa=+$ valley, when the $E_{z}=E_{c}$
(or equivalently, $\alpha=1$), the spin-up band become gapless and forms a Dirac point. Around this critical
electric field ($\alpha=1$), a topological phase transition occurs from the topological insulator phase ($\alpha<1$)
to the band insulator phase ($\alpha>1$). As the vertical electric field further increases ($\alpha>1$), the gapless band reopens a gap.

Compared with other typical monolayer Dirac materials, 1T$^\prime$-MoS$_2$ exhibits several attractive features in
band structure. First, the valley-spin-polarized behaviors are similar to that in silicene \cite{PRBStille2012} except for a band tilting along the $k_y$ direction. Second, the tilted Dirac bands are similar to that of 8-$Pmmn$ borophene \cite{PRBVerma2017} except for a finite SOC gap. Furthermore, the band dispersion along the $k_x$ direction is upright, similar to that in silicene \cite{PRBStille2012}. Third, the band structure of 1T$^\prime$-MoS$_2$ differs strongly from that of its polytypic structure 1H-MoS$_2$ \cite{PRLMoS2010,PRLMoS2012}. These fascinating characteristics make 1T$^\prime$-MoS$_2$ a very interesting monolayer Dirac material. In the following, we will theoretically study the longitudinal optical conductivity in 1T$^\prime$-MoS$_2$.

Within the linear response theory, the longitudinal optical conductivity $\sigma_{jj}(\omega,\alpha)$ at finite
photon frequency $\omega$ is given by
\begin{align}
\sigma_{jj}(\omega)&=\sum_{\kappa=\pm}\sum_{s=\pm}\sigma_{jj}^{\kappa s}(\omega),
\label{totalOC}
\end{align}
where $\sigma_{jj}^{\kappa s}(\omega)$ denotes the longitudinal optical conductivity at given valley $\kappa$
and spin $s$, whose explicit expression can be found in Appendix \ref{Sec:AppendixA}. It is noted that $\sigma_{jj}^{\kappa s}(\omega)$ does not vanish only when the bands contributed to the intraband/interband
transitions share the same spin index (see the Appendix \ref{Sec:AppendixA} for details). To put it equivalently, around the $\kappa$ valley, only when the photon energy $\omega$ is the energy difference between the
band $\xi$ and $\xi^{\prime}$ at given wave vector $\boldsymbol{k}$ and spin index $s$,
\begin{align}
\Delta\varepsilon_{\kappa s}^{\xi\xi^{\prime}}(k,\theta_{\boldsymbol{k}})
&=\varepsilon_{\kappa s}^{\xi^{\prime}}(k_x,k_y)-\varepsilon_{\kappa s}^{\xi}(k_x,k_y)
\nonumber\\&
=(\xi^{\prime}-\xi)\mathcal{Z}_{\kappa s}(k\cos\theta_{\boldsymbol{k}},k\sin\theta_{\boldsymbol{k}}),
\label{Transitionfre}
\end{align}
can an intraband/interband optical transition contribute to the optical conductivity, where $k=|\boldsymbol{k}|$
and $\theta_{\boldsymbol{k}}=\arctan (k_y/k_x)$. It can also be proven that $\sigma_{jj}^{\kappa s}(\omega)
=\sigma_{jj}^{-\kappa -s}(\omega)$ (see Appendix \ref{Sec:AppendixB} for details) such that we are allowed
to focus on either the $\kappa=+$ valley or the $\kappa=-$ valley. For convenience, we restrict our analysis to
the $\kappa=+$ valley hereafter.

After some tedious but straightforward algebra, we express the real part of longitudinal optical conductivity as
\begin{align}
\mathrm{Re}\sigma_{jj}^{\kappa s}(\omega)=\mathrm{Re}\sigma_{jj(\mathrm{intra})}^{\kappa s}(\omega)+\mathrm{Re}\sigma_{jj(\mathrm{inter})}^{\kappa s}(\omega),
\end{align}
where the intraband contribution and interband contribution are given respectively as
\begin{widetext}
\begin{align}
\mathrm{Re}\sigma_{jj(\mathrm{intra})}^{\kappa s}(\omega)
&=\pi\int_{-\infty}^{+\infty}\frac{dk_x}{2\pi}
\int_{-\infty}^{+\infty}\frac{dk_y}{2\pi}
\Theta\left[\mu-\Delta_{\kappa}^{s}(\alpha)\right]
\mathcal{F}_{jj;s,s}^{\kappa;+,+}(k_x,k_y)\left[-\frac{d f\left[\varepsilon_{\kappa s}^{+}(k_x,k_y)\right]}{d \varepsilon_{\kappa s}^{+}(k_x,k_y)}\right]
\delta(\omega)
\nonumber\\&
+\pi\int_{-\infty}^{+\infty}\frac{dk_x}{2\pi}
\int_{-\infty}^{+\infty}\frac{dk_y}{2\pi}
\Theta\left[-\mu-\Delta_{\kappa}^{s}(\alpha)\right]
\mathcal{F}_{jj;s,s}^{\kappa;-,-}(k_x,k_y)\left[-\frac{d f\left[\varepsilon_{\kappa s}^{-}(k_x,k_y)\right]}{d \varepsilon_{\kappa s}^{-}(k_x,k_y)}\right]
\delta(\omega),\\
\mathrm{Re}\sigma_{jj(\mathrm{inter})}^{\kappa s}(\omega)
&=\pi\int_{-\infty}^{+\infty}\frac{dk_x}{2\pi}
\int_{-\infty}^{+\infty}\frac{dk_y}{2\pi}
\mathcal{F}_{jj;s,s}^{\kappa;-,+}(k_x,k_y)
\frac{f[\varepsilon_{\kappa s}^{-}(k_x,k_y)]
-f[\varepsilon_{\kappa s}^{+}(k_x,k_y)]}{\omega}
\delta\left[\omega-2\mathcal{Z}_{\kappa s}(k_x,k_y)\right],
\end{align}
with $\Theta(x)$ the Heaviside step function, $\delta(x)$ the Dirac $\delta$ function, and $f(x)=\{1+\exp{\left[(x
-\mu)/k_B T\right]}\}^{-1}$ the Fermi distribution function in which $\mu$ denotes the chemical potential measured
with respect to the Dirac point, $k_B$ is the Boltzmann constant, and $T$ represents the temperature. It is emphasized
that the physical behaviors of optical conductivity are sensitive to the types of doping which are determined by the criteria in terms of chemical potential $\mu$ and valley-spin-polarized gaps $\Delta_{\kappa}^{s}(\alpha)$ as
\begin{align}
\begin{cases}
$\emph{n}-doped for both $s=+$ and $s=-$$,& \mathrm{Max}\{\Delta_{\kappa}^{-}(\alpha),\Delta_{\kappa}^{+}(\alpha)\}<\mu;\vspace{0.25cm}\\
$\emph{n}-doped for $\kappa s=+$ but undoped for $\kappa s=-$$,& \mathrm{Min}\{\Delta_{\kappa}^{-}(\alpha),\Delta_{\kappa}^{+}(\alpha)\}<
\mu<\mathrm{Max}\{\Delta_{\kappa}^{-}(\alpha),\Delta_{\kappa}^{+}(\alpha)\};\vspace{0.25cm}\\
$undoped for both $s=+$ and $s=-$$,&\mathrm{Max}\{-\Delta_{\kappa}^{-}(\alpha),-\Delta_{\kappa}^{+}(\alpha)\}<
\mu<\mathrm{Min}\{\Delta_{\kappa}^{-}(\alpha),\Delta_{\kappa}^{+}(\alpha)\};\vspace{0.25cm}\\
$\emph{p}-doped for $\kappa s=+$ but undoped for $\kappa s=-$$,& \mathrm{Min}\{-\Delta_{\kappa}^{-}(\alpha),-\Delta_{\kappa}^{+}(\alpha)\}<
\mu<\mathrm{Max}\{-\Delta_{\kappa}^{-}(\alpha),-\Delta_{\kappa}^{+}(\alpha)\};\vspace{0.25cm}\\
$\emph{p}-doped  for both $s=+$ and $s=-$$,& \mu<\mathrm{Min}\{-\Delta_{\kappa}^{-}(\alpha),-\Delta_{\kappa}^{+}(\alpha)\}.
\end{cases}
\label{dopingcriteria}
\end{align}
\end{widetext}
Especially, in the absence of electric field ($\alpha=0$), the energy bands are spin-degenerate and the energy
gaps $\Delta_{\kappa}^{s}(0)=\Delta_{\mathrm{so}}$ are valley- and spin-independent. It is undoped when $-\Delta_{\mathrm{so}}<\mu<\Delta_{\mathrm{so}}$ where the valence bands are fully occupied but the conduction
bands are empty, $n$-doped when $\mu>\Delta_{\mathrm{so}}$ where the conduction bands
are partially filled with electrons, and $p$-doped when $\mu<-\Delta_{\mathrm{so}}$ where the valence bands are
partially filled with electrons. It is further pointed out that the zero-frequency part due to the intraband transitions contributes to the Drude peak, whereas the frequency-dependent part is associated with the interband transitions between a filled band to an empty band.

\section{Numerical results\label{Sec:NumericalResults}}

In this section, we show the numerical results for the absorption peaks of longitudinal conductivities in the
undoped and doped 1T$^\prime$-MoS$_2$ when the vertical electric field is absent and present, respectively. We
set $T=1\mathrm{K}$ throughout the numerical calculation in this paper.

\subsection{In the absence of vertical electric field}

\begin{figure}[htbp]
\centering
\includegraphics[width=8cm]{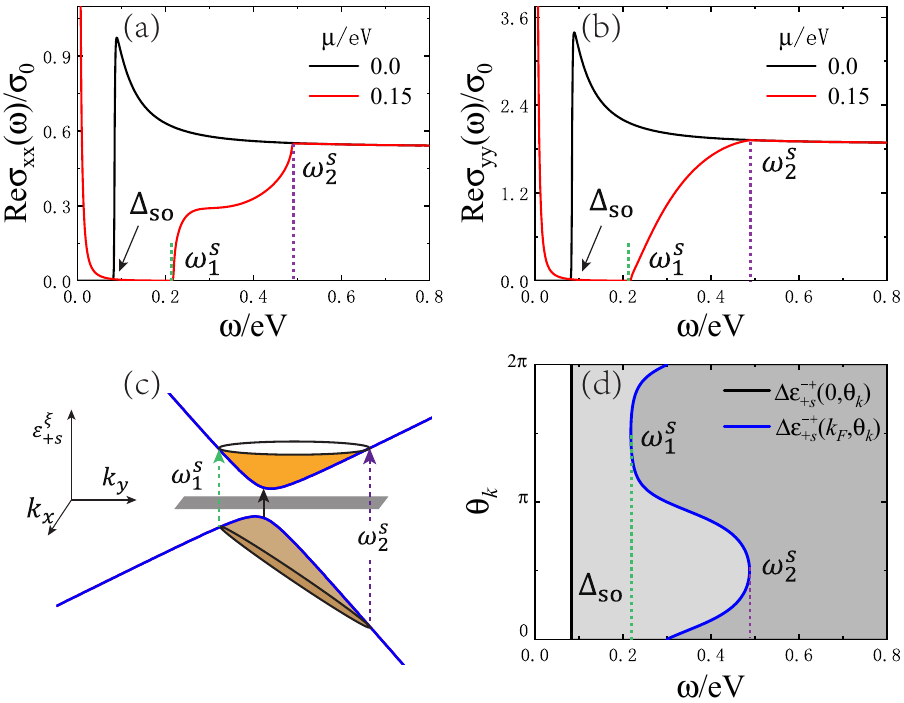}
\caption{(Color online) Total longitudinal optical conductivities of 1T$^\prime$-MoS$_2$ in the absence of vertical electric field. In (a) and (b), the black lines represent the optical conductivity in the undoped case in which the absorption peak is at $\omega=\Delta_{\mathrm{so}}$ while the red lines stand for the optical conductivity in the doped case where two spin-degenerate absorption peaks are at $\omega=\omega_{1}^{s}$ and $\omega=\omega_{2}^{s}$. Panel (c) shows the schematic diagram of interband transition for the $\kappa=+$ valley, where the colored arrows and black arrows indicate the boundary of electron transition for $n$-doped case and undoped case, respectively. The blue curve (for doped case) and black line (for undoped case) in the panel (d) show the lower boundaries of incident photon frequency $\omega$ for arbitrary direction of wavevector $\theta_{\boldsymbol{k}}$ by using Eq.(\ref{Transitionfre}). Note that $\theta_{\boldsymbol{k}}=\pi/2$ and $\theta_{\boldsymbol{k}}=3\pi/2$ correspond to $\omega_{2}^{s}$ and $\omega_{1}^{s}$, respectively.}
\label{fig2}
\end{figure}

Without applying the vertical electric field ($\alpha=0$), the energy bands of 1T$^\prime$-MoS$_2$ are fully gapped
and spin-degenerate such that the total optical conductivity $\mathrm{Re}\sigma_{jj}(\omega)=g_vg_s\mathrm{Re}
\sigma_{jj}^{+s}(\omega)$ with $g_v$ and $g_s$ the degeneracy factors of valley and spin. As shown in Fig.(\ref{fig2}), $\mathrm{Re}\sigma_{xx}(\omega)$ and $\mathrm{Re}\sigma_{yy}(\omega)$ exhibit a strong anisotropy because the
dispersion is tilted along the $k_y$ direction but untilted along the $k_x$ direction, similar to that in 8-$Pmmn$ borophene \cite{PRBVerma2017} but unlike that in 1H-MoS$_2$ \cite{PRBCarbotte2012,PRBCarvalho2013,PRBAsgari2014}. When we set $v_{-}=v_{t}$, $v_{2}=0$, $v_{1}=v_{x}$, $v_{+}=v_{y}$, and $\Delta_{\mathrm{so}}=0$, our results go back to that in 8-$Pmmn$ borophene \cite{PRBVerma2017}, and when we set $v_{-}=v_{2}=0$ and $v_{+}=v_{1}=v_{F}$, our results restore that in silicene \cite{PRBStille2012}.

Specifically, for the undoped case ($-\Delta_{\mathrm{so}}<\mu<\Delta_{\mathrm{so}}$), the optical conductivity
is contributed entirely by the interband transitions, which can be seen from Fig.(\ref{fig2}c). As a consequence, $\mathrm{Re}\sigma_{xx}(\omega)$ and $\mathrm{Re}\sigma_{yy}(\omega)$ share the same peak that appears
at the absorption edge where $\omega=\Delta_{\mathrm{so}}$, as shown in Fig.(\ref{fig2}a)
and Fig.(\ref{fig2}b). Hence, from this peak in experimental measurement the SOC gap $\Delta_{\mathrm{so}}$ can
be precisely determined. From Fig.(\ref{fig2}a) and Fig.(\ref{fig2}b), $\mathrm{Re}\sigma_{xx}(\omega)$ and $\mathrm{Re}\sigma_{yy}(\omega)$ decay respectively to different asymptotic background values
\begin{align}
\mathrm{Re}\sigma_{xx(\mathrm{asyp})}&=\frac{v_1}{\sqrt{v_2^2+v_{+}^2}}\sigma_0,\\
\mathrm{Re}\sigma_{yy(\mathrm{asyp})}&=\frac{\sqrt{v_2^2+v_{+}^2}}{v_1}\sigma_0,
\end{align}
in the regime of large photon energy, where $\sigma_{0}=e^2/4\hbar$ (we restore $\hbar$ for explicitness). It is
noted that the different values originate from the strong anisotropy of Fermi velocities $v_1$ and $\sqrt{v_2^2+v_{+}^2}$ instead of the tilting parameterized by $v_{-}$, and that the asymptotic values result from
the asymptotic linearity of Dirac band in the high-energy regime. These properties can be well understood by the analytical results of the gapless model ($\Delta_{\mathrm{so}}=0$) in Appendix \ref{Sec:AppendixC}. In addition, these two asymptotic values satisfy a universal relation $\mathrm{Re}\sigma_{xx(\mathrm{asyp})}\times\mathrm{Re}\sigma_{yy(\mathrm{asyp})}
=\sigma_{0}^2$. It is emphasized that this kind of universal relation holds not only for the undoped case presented above but also for the doped case to be shown below (see Appendix \ref{Sec:AppendixC} for more details and explanations), which has also been reported in other tilted Dirac systems, such as the tilted gapless Dirac bands
in 8-$Pmmn$ borophene \cite{PRBVerma2017}.

In the $n$-doped case ($\mu>\Delta_{\mathrm{so}}$), the optical conductivities $\mathrm{Re}\sigma_{xx}(\omega)$ and $\mathrm{Re}\sigma_{yy}(\omega)$ exhibit more interesting physics, as shown in Fig.(\ref{fig2}a) and Fig.(\ref{fig2}b). First, both $\mathrm{Re}\sigma_{xx}(\omega)$ and $\mathrm{Re}\sigma_{yy}(\omega)$ possess a Drude peak around $\omega=0$ due to the intraband transitions. Second, they share two absorption peaks depending on the chemical potential at $\omega=\omega_{1}^{s}$ and $\omega=\omega_{2}^{s}$ due to the band tilting along the $k_y$ direction.
Third, the Pauli blocking prevents the optical transition between valence band and conduction band entirely when $0<\omega<\omega_{1}^{s}$ but partially when $\omega_{1}^{s}<\omega<\omega_{2}^{s}$. Fourth, when $\omega>\omega_{2}^{s}$, $\mathrm{Re}\sigma_{xx}(\omega)$ and $\mathrm{Re}\sigma_{yy}(\omega)$ approach respectively to their asymptotic background values $\mathrm{Re}\sigma_{xx(\mathrm{asyp})}$ and $\mathrm{Re}\sigma_{yy(\mathrm{asyp})}$, due to the linearity of Dirac band in the high-energy regime. All these interesting behaviors can also be directly
read out from either Fig.(\ref{fig2}c) or Fig.(\ref{fig2}d). It is further remarked that for the $p$-doped case ($\mu<-\Delta_{\mathrm{so}}$), there is no qualitative difference in the behaviors of longitudinal optical conductivity presented here.

\subsection{In the presence of vertical electric field}

\begin{figure}[htbp]
\centering
\includegraphics[width=8cm]{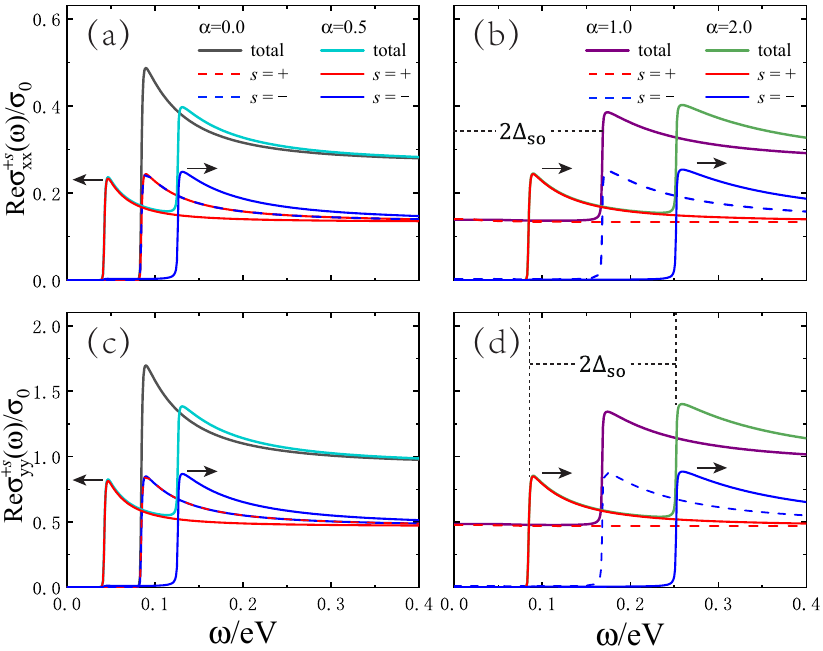}
\caption{(Color online) Valley-spin-polarized optical conductivities $\mathrm{Re}\sigma_{xx}^{\kappa s}(\omega)$ and $\mathrm{Re}\sigma_{yy}^{\kappa s}(\omega)$ in the undoped 1T$^\prime$-MoS$_2$ when the vertical electric field is present. The arrows indicate the moving tendency of peaks as the electric field smoothly increases. The chemical potential is set to be $\mu=0$ which corresponds to the undoped case ($-\Delta_{\mathrm{so}}<\mu<\Delta_{\mathrm{so}}$) when $\alpha=0$. We take the case for the $\kappa=+$ valley as a demonstration.}
\label{fig3}
\end{figure}

\begin{figure}[htbp]
\centering
\includegraphics[width=8cm]{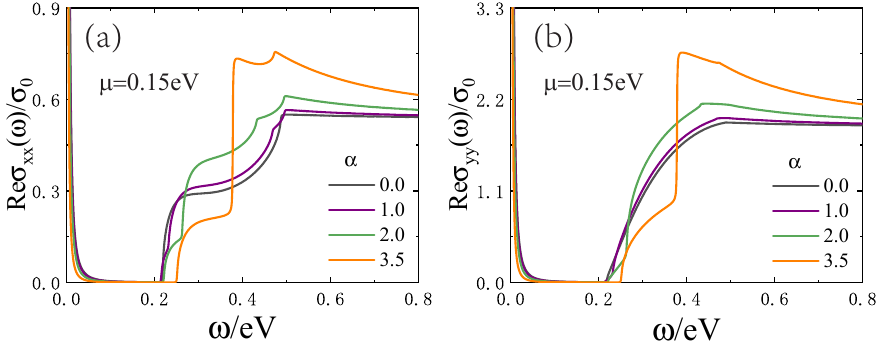}
\caption{(Color online) Dependence of total optical conductivities $\mathrm{Re}\sigma_{xx}$ and $\mathrm{Re}\sigma_{yy}$ on the vertical electric field in 1T$^\prime$-MoS$_2$. The chemical potential is set to be $\mu=0.15~\mathrm{eV}$ which corresponds to the $n$-doped case ($\mu>\Delta_{\mathrm{so}}$) when $\alpha=0$.}
\label{fig4}
\end{figure}

\begin{figure*}[tbp]
\centering
\includegraphics[width=16cm]{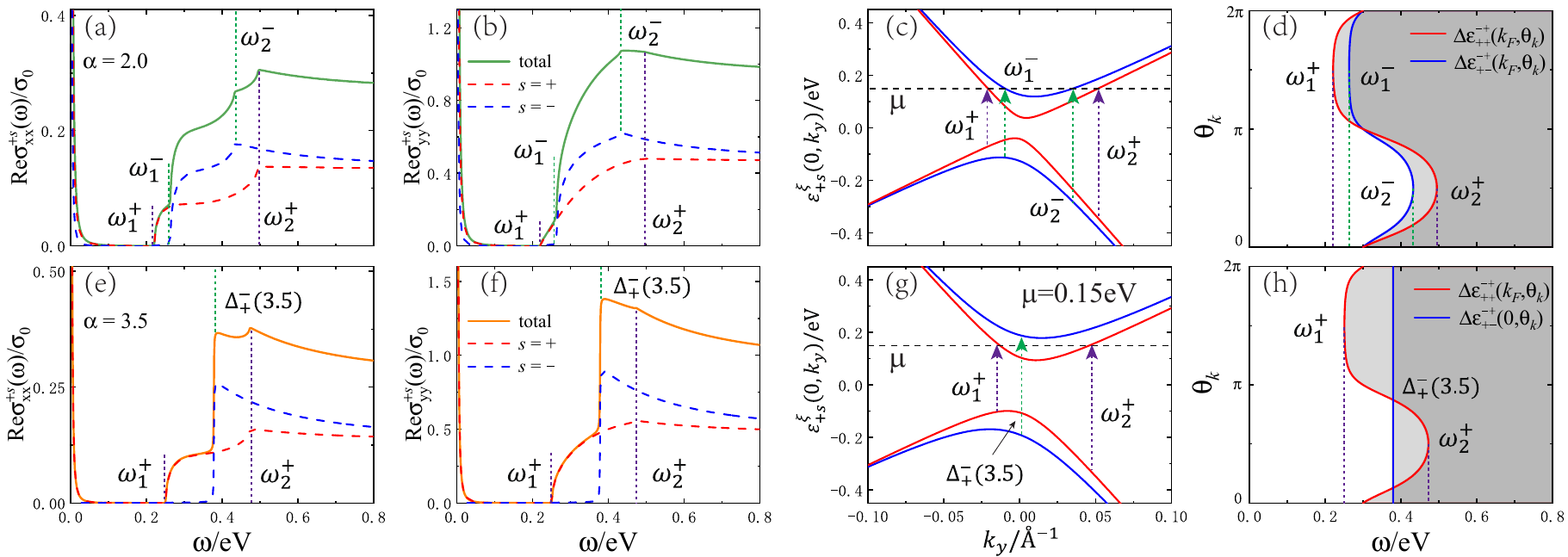}
\caption{(Color online) Valley-spin-polarized longitudinal conductivities $\mathrm{Re}\sigma_{xx}^{\kappa s}(\omega)$ and $\mathrm{Re}\sigma_{yy}^{\kappa s}(\omega)$ in 1T$^\prime$-MoS$_2$ for two specific values of vertical electric field. The chemical potential is set to be $\mu=0.15~\mathrm{eV}$ which corresponds to the $n$-doped case ($\mu>\Delta_{\mathrm{so}}$) when $\alpha=0$. We take the case for the $\kappa=+$ valley as a demonstration.}
\label{fig5}
\end{figure*}

When the vertical electric field is applied ($\alpha>0$), the valley-spin-polarized bands and gaps of 1T$^\prime$-MoS$_2$ result in many interesting changes in the longitudinal optical conductivity. As shown in Figs.(\ref{fig3}-\ref{fig5}), $\mathrm{Re}\sigma_{xx}^{+s}(\omega)$ and $\mathrm{Re}\sigma_{yy}^{+s}(\omega)$
are generally anisotropic and valley-spin-polarized, which is strongly different from that in 1H-MoS$_2$ \cite{PRBCarbotte2012,PRBCarvalho2013,PRBAsgari2014}. When we set $v_{-}=v_{2}=0$ and $v_{+}=v_{1}=v_{F}$, our
results restore that in silicene \cite{PRBStille2012}.

For $-\Delta_{\mathrm{so}}<\mu<\Delta_{\mathrm{so}}$ which corresponds to the undoped case when $\alpha=0$, the vertical electric field splits the spin-degenerate absorption edge $\omega=\Delta_{+}^{s}(0)=\Delta_{\mathrm{so}}$
into two spin-polarized absorption edges $\omega=\Delta_{+}^{+}(\alpha)$ and $\omega=\Delta_{+}^{-}(\alpha)$ as shown in Fig.(\ref{fig3}), where the dependence of valley-spin-polarized gaps $\Delta_{\kappa}^{s}(\alpha)=|\kappa-\alpha s|\Delta_{\mathrm{so}}$ on the electric field was previously shown in Fig.(\ref{fig1}). In the region of topological insulator phase ($0\le\alpha<1$), two spin-polarized peaks $\omega=\Delta_{+}^{+}(\alpha)$ and $\omega=\Delta_{+}^{-}(\alpha)$ for the $\kappa=+$ valley move oppositely as the electric field smoothly changes,
which can be seen in Fig.(\ref{fig3}a) and Fig.(\ref{fig3}c). However, in the region of band insulator phase ($\alpha>1$), two spin-polarized peaks $\omega=\Delta_{+}^{+}(\alpha)$ and $\omega=\Delta_{+}^{-}(\alpha)$ separated
by $2\Delta_{\mathrm{so}}$ shift in concert towards the higher frequency as shown in Fig.(\ref{fig3}b) and Fig.(\ref{fig3}d). Interestingly, at the critical electric field ($\alpha=1$), the gapped band remains the larger absorption edge at $\omega=\Delta_{+}^{+}(1)$ while the gapless band vanishes as the smaller absorption edge at $\omega=\Delta_{+}^{-}(1)=0$. Moreover, when $0<\omega<\Delta_{+}^{+}(1)$, the value of the step represented by the black line is half of the asymptotic value, indicating that only the gapless band contributes to the optical transition, whereas when $\omega>\Delta_{+}^{+}(1)$, the gapped band also contributes to the optical transition such that the optical conductivity approaches the asymptotic values in the large-$\omega$ regime. The total optical conductivity is obtained by summing over valley
index $\kappa$ and spin index $s$, leading to a superposition of four valley-spin-polarized optical conductivities $\mathrm{Re}\sigma_{jj}^{\kappa s}$. The exotic behaviors of absorption edges can be used to probe the topological phase transition induced by the vertical electric field, which has also been reported in other similar systems, such
as the untilted gapped Dirac bands in silicene \cite{PRBStille2012}.

For $\mu>\Delta_{\mathrm{so}}$ which corresponds to the $n$-doped case when $\alpha=0$, the total optical conductivities exhibit many richer features after the vertical electric field is applied. It can be
found from Fig.(\ref{fig4}a) that for different vertical electric fields $\mathrm{Re}\sigma_{xx}(\omega)$
share the same asymptotic values in the regime of large photon energy and that the absorption peaks change
with the electric field, which also holds for $\mathrm{Re}\sigma_{yy}(\omega)$ as shown in Fig.(\ref{fig4}b). Specifically, we show two characteristic cases as a demonstration in Fig.(\ref{fig5}). From the upper panels
of Fig.(\ref{fig5}), when $\alpha=2$, the Fermi level cuts both spin-up and spin-down conduction bands, which
indicates that both spin-polarized energy bands are $n$-doped according to the criteria in (\ref{dopingcriteria}).
Therefore two spin-degenerate peaks $\omega=\omega_{1}^{s}$ and $\omega=\omega_{2}^{s}$ split further into
four spin-polarized peaks $\omega=\omega_{1}^{+}$, $\omega=\omega_{1}^{-}$, $\omega=\omega_{2}^{+}$, and $\omega=\omega_{2}^{-}$. However, from the lower panels of Fig.(\ref{fig5}), when $\alpha=3.5$, the Fermi level
cuts the spin-up conduction band but lies in the gap of spin-down bands, which indicates that one kind of spin-polarized energy band is $n$-doped but the other kind of spin-polarized energy band is undoped according to the criteria in (\ref{dopingcriteria}). As a consequence, $\omega=\omega_{1}^{s}$ split further into two spin-polarized peaks $\omega=\omega_{1}^{+}$ and $\omega=\omega_{1}^{-}$, but $\omega=\Delta_{+}^{-}(3.5)$ does not split and hence contributes only to one peak. These lead to an interesting dependence of total optical conductivities on the magnitude of the vertical electric field. Therefore, one can select some specific frequencies of absorbed photons by tuning the vertical electric field. And all of these behaviors can also be directly read out from Figs.(\ref{fig5}c), (\ref{fig5}d), (\ref{fig5}g), and (\ref{fig5}h). It is further remarked that for $\mu<-\Delta_{\mathrm{so}}$ which corresponds to the $p$-doped case when $\alpha=0$, the behaviors of longitudinal optical conductivities do not differ qualitatively from those presented here.

\section{Summary and Conclusions\label{Sec:Conclusions}}

In this work, we have theoretically studied the anisotropic longitudinal optical conductivities of tilted Dirac
bands in both undoped and doped 1T$^\prime$-MoS$_2$, including effects of an external vertical electric field. In the absence of vertical electric field, the tilted Dirac bands are spin-degenerate. For the undoped 1T$^\prime$-MoS$_2$, the longitudinal conductivities are contributed entirely by the interband transitions and share the same peak at the absorption edge, from which the SOC gap can be directly extracted from experimental measurements. The longitudinal optical conductivities decay respectively to different asymptotic background values originated from the anisotropy and the asymptotic linearity of Dirac band. For the doped 1T$^\prime$-MoS$_2$, two longitudinal conductivities possess a Drude peak around $\omega=0$ due to the intraband transitions and share two absorption peaks depending on the chemical potential at $\omega=\omega_{1}^{s}$ and $\omega=\omega_{2}^{s}$ due to the band tilting along the $k_y$ direction. The Pauli-blocking prevents the optical transition between the valence band and conduction band entirely when $0<\omega<\omega_{1}^{s}$ but partially
when $\omega_{1}^{s}<\omega<\omega_{2}^{s}$. When $\omega>\omega_{2}^{s}$, two longitudinal conductivities approach respectively to their asymptotic background values due to the linearity of Dirac band in the high-energy regime.
In general, for both undoped and doped 1T$^\prime$-MoS$_2$, the asymptotic background values always satisfy a
universal relation $\mathrm{Re}\sigma_{xx(\mathrm{asyp})}\times\mathrm{Re}\sigma_{yy(\mathrm{asyp})}=\sigma_{0}^2$
with $\sigma_{0}=e^2/4\hbar$.

In the presence of vertical electric field, the tilted Dirac bands in 1T$^\prime$-MoS$_2$ are valley-spin-polarized
due to the interplay between SOC gap and electric field, and hence a topological phase transition occurs between
the topological insulator phase and the band insulator phase with a valley-spin-polarized metal state at a critical
value in between. For the undoped 1T$^\prime$-MoS$_2$, when the vertical electric field increases smoothly, the interband absorption edge splits into two absorption edges marked by two peaks of longitudinal optical conductivities. These two peaks move oppositely in frequency in the region of the topological insulator phase and shift in concert
to a higher frequency in the region of the band insulator phase. The exotic moving of optical conductivity due to
the vertical electric field can be taken as a fingerprint of topological phase transition in the undoped 1T$^\prime$
-MoS$_2$. For the doped 1T$^\prime$-MoS$_2$, the absorption peaks of longitudinal optical conductivity change with respect to the competition among the SOC gap, electric field, and chemical potential. By manipulating the vertical electric field, one can select the specific frequencies of absorbed photons.

In conclusion, the influence of SOC gap, band tilting, and external vertical electric field on the anisotropic longitudinal optical conductivities of tilted Dirac bands in 1T$^\prime$-MoS$_2$ was revealed. A theoretical scheme
for probing the topological phase transition in 1T$^\prime$-MoS$_2$ via exotic behaviors of longitudinal optical conductivities was proposed. This work provided critical insights into understanding the tilted Dirac bands of 1T$^\prime$ TMDCs in general, and their contributions to optical properties of 1T$^\prime$-MoS$_2$ in particular.
The results for 1T$^\prime$-MoS$_2$ are expected to be qualitatively valid for other monolayer tilted gapped Dirac materials, such as $\alpha$-SnS$_2$, TaCoTe$_2$, and TaIrTe$_4$, due to the similarity in their band structures.

\section*{ACKNOWLEDGEMENTS\label{Sec:acknowledgements}}

We are grateful to Xiang-Hua Kong, Yuanpei Lan, and Zhen-Bing Dai for valuable discussions. This work is partially supported by the National Natural Science Foundation of China under Grants No.11547200 and No.11874273, the China Scholarship Council (Grant No.201608515061), and the Key Project of Sichuan Science and Technology Program (2019YFSY0044). H.G. and H.-R.C. were also supported by the NSERC of Canada and the FQRNT of Quebec (H.G.). We thank the High Performance Computing Centers at Sichuan Normal University,  McGill University, and Compute Canada.

\appendix

\begin{widetext}

\section{Explicit expression of the longitudinal optical conductivity\label{Sec:AppendixA}}

Within the linear response theory, the longitudinal optical conductivity at the $\kappa$ valley can
be expressed as
\begin{align}
\sigma_{jj}^{\kappa}(\omega)&=\frac{i}{\omega}
\frac{1}{\beta_{T}}\sum_{\Omega_m}\int_{-\infty}^{+\infty}\frac{dk_x}{2\pi}\int_{-\infty}^{+\infty}\frac{dk_y}{2\pi}
\mathrm{Tr}\left[\hat{J}_{j}^{\kappa}G_{\kappa}(k_x,k_y,i\Omega_m)
\hat{J}_{j}^{\kappa}G_{\kappa}(k_x,k_y,i\Omega_m+\omega+i\eta)\right],
\end{align}
where $\beta_{T}=1/k_B T$, $j=x,y$ refer to spatial coordinates, $\eta$ denotes a positive infinitesimal. The charge
current operators read
\begin{align}
\hat{J}_x^{\kappa}&=e\frac{\partial H_\kappa(k_x,k_y)}{\partial k_x}=ev_1\gamma_1,\\
\hat{J}_y^{\kappa}&=e\frac{\partial H_\kappa(k_x,k_y)}{\partial k_y}
=e\left(v_2\gamma_0-\kappa v_{-}I-\kappa v_{+}\gamma_2\right),
\end{align}
and the Matsubara Green's function takes the form
\begin{align}
G_{\kappa}(k_x,k_y,i\Omega_m)&
=\left[i\Omega_m+\mu-H_\kappa(k_x,k_y)\right]^{-1}
=\frac{1}{4}\sum_{s,\xi=\pm}\frac{\mathcal{P}_{\kappa s}^{\xi}(k_x,k_y)}
{i\Omega_m+\mu-\varepsilon_{\kappa s}^{\xi}(k_x,k_y)},
\end{align}
where $\mu$ is the chemical potential, and
\begin{align}
\mathcal{P}_{\kappa s}^{\xi}(k_x,k_y)
&=\left[\tau_0+\xi\frac{-s A_{\kappa s}(k_y)\tau_1+B(k_x)\tau_2
+C_{\kappa}(k_y)\tau_3}{\mathcal{Z}_{\kappa s}(k_x,k_y)}\right]
\otimes\left(\sigma_0-s \sigma_1\right),\\
\varepsilon_{\kappa s}^{\xi}(k_x,k_y)
&=-\kappa v_{-}k_y+ \xi\mathcal{Z}_{\kappa s}(k_x,k_y),\label{EnergyDef}
\end{align}
with
\begin{align}
A_{\kappa s}(k_y)&=v_2k_y+(\kappa-\alpha s)\Delta_{\mathrm{so}},\label{ADef}\\
B(k_x)&=v_{1}k_x,\label{BDef}\\
C_{\kappa}(k_y)&=-\kappa v_{+}k_y,\\
\mathcal{Z}_{\kappa s}(k_x,k_y)
&=\sqrt{\left[A_{\kappa s}(k_y)\right]^2+\left[B(k_x)\right]^2+\left[C_{\kappa}(k_y)\right]^2}.\label{CDef}
\end{align}

After summing over Matsubara frequency $\Omega_m$, we express the longitudinal optical conductivity at the
$\kappa$ valley as
\begin{align}
\sigma_{jj}^{\kappa}(\omega)&
=\frac{i}{\omega}\int_{-\infty}^{+\infty}\frac{dk_x}{2\pi}\int_{-\infty}^{+\infty}\frac{dk_y}{2\pi}
\sum_{s,s^{\prime}=\pm}\sum_{\xi,\xi^{\prime}=\pm}
\mathcal{F}_{jj;s,s^{\prime}}^{\kappa;\xi,\xi^{\prime}}(k_x,k_y)
\mathcal{M}_{s,s^{\prime}}^{\kappa;\xi,\xi^{\prime}}(k_x,k_y,\omega),
\label{DefOCApp}
\end{align}
where
\begin{align}
\mathcal{F}_{jj;s,s^{\prime}}^{\kappa;\xi,\xi^{\prime}}(k_x,k_y)
=\frac{\mathrm{Tr}[\hat{J}_j^{\kappa}\mathcal{P}_{\kappa s}^{\xi}(k_x,k_y)
\hat{J}_j^{\kappa}\mathcal{P}_{\kappa s^{\prime}}^{\xi^{\prime}}(k_x,k_y)]}{16},
\end{align}
and
\begin{align}
\mathcal{M}_{s,s^{\prime}}^{\kappa;\xi,\xi^{\prime}}(k_x,k_y,\omega)
&=\frac{f[\varepsilon_{\kappa s}^{\xi}(k_x,k_y)]
-f[\varepsilon_{\kappa s^{\prime}}^{\xi^{\prime}}(k_x,k_y)]} {\omega+\varepsilon_{\kappa s}^{\xi}(k_x,k_y)
-\varepsilon_{\kappa s^{\prime}}^{\xi^{\prime}}(k_x,k_y)+i\eta},\label{MDef}
\end{align}
with $f(x)=1/\left\{1+\exp[(x-\mu)/(k_B T)]\right\}$ denoting the Fermi distribution function.

Specifically, the explicit expressions of $\mathcal{F}_{jj;s,s^{\prime}}^{\kappa;\xi,\xi^{\prime}}(k_x,k_y)$
are given as
\begin{align}
\mathcal{F}_{xx;s,s^{\prime}}^{\kappa;\xi,\xi^{\prime}}(k_x,k_y)
&=\frac{\mathrm{Tr}[\hat{J}_x^{\kappa}\mathcal{P}_{\kappa s}^{\xi}(k_x,k_y)
\hat{J}_x^{\kappa}\mathcal{P}_{\kappa s^{\prime}}^{\xi^{\prime}}(k_x,k_y)]}{16}
=\frac{e^2v_1^2}{2}\delta_{s s^{\prime}}\left\{1-\xi\xi^{\prime}\frac{\left[A_{\kappa s}(k_y)\right]^2
-\left[B(k_x)\right]^2
+\left[C_{\kappa}(k_y)\right]^2}{\left[\mathcal{Z}_{\kappa s}(k_x,k_y)\right]^2}
\right\},\label{FxxDef}
\end{align}
and
\begin{align}
\hspace{-0.5cm}\mathcal{F}_{yy;s,s^{\prime}}^{\kappa;\xi,\xi^{\prime}}(k_x,k_y)
&=\frac{\mathrm{Tr}[\hat{J}_y^{\kappa}\mathcal{P}_{\kappa s}^{\xi}(k_x,k_y)
\hat{J}_y^{\kappa}\mathcal{P}_{\kappa s^{\prime}}^{\xi^{\prime}}(k_x,k_y)]}{16}
\nonumber\\&
=\frac{e^2}{2}\delta_{ss^{\prime}}\left\{\left(v_2^2+v_{-}^2+v_{+}^2\right)
-\xi\xi^{\prime}\kappa v_2v_{+}\frac{4 A_{\kappa s}(k_y)C_{\kappa}(k_y)}{\left[\mathcal{Z}_{\kappa s}(k_x,k_y)\right]^2}
+2(\xi+\xi^{\prime})v_{-}\frac{v_{+}C_{\kappa}(k_y)-\kappa v_2A_{\kappa s}(k_y)}{\mathcal{Z}_{\kappa s}(k_x,k_y)}
\right.\nonumber\\&\left.\hspace{0.5cm}
+\xi\xi^{\prime}\frac{\left(v_2^2+v_{-}^2-v_{+}^2\right)\left[A_{\kappa s}(k_y)\right]^2
-\left(v_2^2-v_{-}^2+v_{+}^2\right)\left[B(k_x)\right]^2-
\left(v_2^2-v_{-}^2-v_{+}^2\right)\left[C_{\kappa}(k_y)\right]^2}{\left[\mathcal{Z}_{\kappa s}(k_x,k_y)\right]^2}
\right\}.\label{FyyDef}
\end{align}
It is noted that $\delta_{ss^{\prime}}$ in Eqs. (\ref{FxxDef}) and (\ref{FyyDef}) implies that both $\mathcal{F}_{xx;s,s^{\prime}}^{\kappa;\xi,\xi^{\prime}}(k_x,k_y)$ and $\mathcal{F}_{yy;s,s^{\prime}}^{\kappa;\xi,\xi^{\prime}}(k_x,k_y)$ vanish when $s^{\prime}\neq s$, which indicates
that the bands contributed to the intraband/interband transitions must share the same spin index.

Therefore, the longitudinal optical conductivity at the $\kappa$ valley can be expressed as
\begin{align}
\sigma_{jj}^{\kappa}(\omega)&=\sum_{s=\pm}\sigma_{jj}^{\kappa s}(\omega),
\end{align}
where
\begin{align}
\sigma_{jj}^{\kappa s}(\omega)&
=\frac{i}{\omega}\int_{-\infty}^{+\infty}\frac{dk_x}{2\pi}\int_{-\infty}^{+\infty}\frac{dk_y}{2\pi}
\sum_{\xi,\xi^{\prime}=\pm}
\mathcal{F}_{jj;s,s}^{\kappa;\xi,\xi^{\prime}}(k_x,k_y)
\mathcal{M}_{s,s}^{\kappa;\xi,\xi^{\prime}}(k_x,k_y,\omega)
\label{VSPOC}
\end{align}
denotes the valley-spin-polarized complex longitudinal conductivity at finite photon frequency $\omega$.

\section{Relation between the valley-spin-polarized longitudinal conductivities with opposite valley and spin \label{Sec:AppendixB}}

From the definitions in Eq.(\ref{EnergyDef}) and Eqs.(\ref{ADef}-\ref{CDef}), we have the following relations
\begin{align}
A_{-\kappa -s}(-k_y)&=-A_{\kappa s}(k_y),\\
B(k_x)&=B(k_x),\\
C_{-\kappa}(-k_y)&=C_{\kappa}(k_y),\\
\mathcal{Z}_{-\kappa -s}(k_x,-k_y)&=\mathcal{Z}_{\kappa s}(k_x,k_y),\\
\varepsilon_{-\kappa -s}^\xi(k_x,-k_y)&=\varepsilon_{\kappa s}^\xi(k_x,k_y).
\end{align}

Substituting these relations into Eqs.(\ref{MDef}-\ref{FyyDef}), we consequently get
\begin{align}
\mathcal{M}_{-s,-s}^{-\kappa;\xi\xi^\prime}(k_x,-k_y,\omega)
&=\mathcal{M}_{s,s}^{\kappa;\xi\xi^\prime}(k_x,k_y,\omega),\label{MRELATION}\\
\mathcal{F}_{xx;-s,-s}^{-\kappa;\xi\xi^\prime}(k_x,-k_y)
&=\mathcal{F}_{xx;s,s}^{\kappa;\xi\xi^\prime}(k_x,k_y),\label{FXXRELATION}\\
\mathcal{F}_{yy;-s,-s}^{-\kappa;\xi\xi^\prime}(k_x,-k_y)
&=\mathcal{F}_{yy;s,s}^{\kappa;\xi\xi^\prime}(k_x,k_y).\label{FYYRELATION}
\end{align}

After substituting these relations into Eq.(\ref{VSPOC}), we obtain the relation
\begin{align}
\sigma_{jj}^{-\kappa -s}(\omega)=\frac{i}{\omega} \int_{-\infty}^{+\infty}\frac{dk_x}{2\pi} \int_{-\infty}^{+\infty}\frac{dk_y}{2\pi}
\sum_{\xi,\xi^\prime=\pm}\mathcal{F}_{jj;-s,-s}^{-\kappa;\xi\xi^\prime}(k_x,-k_y)
\mathcal{M}_{-s,-s}^{-\kappa;\xi\xi^\prime}(k_x,-k_y,\omega)=\sigma_{jj}^{\kappa s}(\omega).
\end{align}

\section{Analytical results in the gapless model and asymptotic background values\label{Sec:AppendixC}}

The asymptotic values of longitudinal optical conductivities denote the values contributed completely by the interband transition in the high-energy regime where the energy of incident photon $\omega$ is asymptotically infinite, which
is equivalent to that the energy of electron $\varepsilon_{\kappa s}^{\xi}(k_x,k_y)$ is asymptotically infinite from the definition $\omega=\varepsilon_{\kappa s}^{\xi^{\prime}}(k_x,k_y)-\varepsilon_{\kappa s}^{\xi}(k_x,k_y)$. In this regime, the influence of the band gap can be neglected such that the gapped model Hamiltonian reduces to its gapless counterpart whose eigenvalues are given as
\begin{align}
\varepsilon_{\kappa s}^{\xi}(k_x,k_y)
&=-\kappa v_t k_y+ \xi\sqrt{v_{x}^2k_x^2+v_y^2k_y^2},
\end{align}
where $v_x=v_1$, $v_y=\sqrt{v_2^2+v_{+}^2}$, and $v_t=v_{-}$. It is noted that they are the same as the energy bands
of 8-$Pmmn$ borophene. In the gapless model, the real part of the longitudinal optical conductivity is given by
\begin{align}
\mathrm{Re}\sigma_{jj}(\omega)&=\sum_{\kappa=\pm}\mathrm{Re}\sigma_{jj}^{\kappa}(\omega)
=\sum_{\kappa=\pm}\sum_{s=\pm}\mathrm{Re}\sigma_{jj}^{\kappa s}(\omega).
\end{align}

To make the following calculation more general, instead of focusing on the asymptotic background values at $\omega\to\infty$, we proceed with the longitudinal optical conductivity for arbitrary positive $\omega$, which is contributed completely by the interband transition. The real part of the interband longitudinal optical conductivity for valley $\kappa$ and spin $s$ reads
\begin{align}
\mathrm{Re}\sigma_{jj(\mathrm{inter})}^{\kappa s}(\omega)
&=\pi\int_{-\infty}^{+\infty}\frac{dk_x}{2\pi}
\int_{-\infty}^{+\infty}\frac{dk_y}{2\pi}
\sum_{s^\prime=\pm}\mathcal{F}_{jj;ss^\prime}^{\kappa;-,+}(k_x,k_y)
\frac{f[\varepsilon_{\kappa s}^{-}(k_x,k_y)]
-f[\varepsilon_{\kappa s^\prime}^{+}(k_x,k_y)]}{\omega}
\delta\left[\omega-2\sqrt{v_{x}^2k_x^2+v_y^2k_y^2}\right],
\end{align}
where $\mathcal{F}_{xx;ss^\prime}^{\kappa;-,+}(k_x,k_y)$ and $\mathcal{F}_{yy;ss^\prime}^{\kappa;-,+}(k_x,k_y)$ are given as
\begin{align}
\mathcal{F}_{xx;ss^\prime}^{\kappa;-,+}(k_x,k_y) &=4\sigma_0\frac{v_x^2v_y^2k_y^2}{v_x^2k_x^2+v_{y}^2k_y^2}\delta_{ss^\prime},\\
\mathcal{F}_{yy;ss^\prime}^{\kappa;-,+}(k_x,k_y) &=4\sigma_0\frac{v_y^2v_x^2k_x^2}{v_x^2k_x^2+v_{y}^2k_y^2}\delta_{ss^\prime}.
\end{align}

After introducing $\tilde{k}_x=v_xk_x$, $\tilde{k}_y=v_yk_y$, $\tilde{k}=|\tilde{\boldsymbol{k}}|=\sqrt{\tilde{k}_x^2+\tilde{k}_y^2}$, $\phi=\arctan (\tilde{k}_y/\tilde{k}_x)$, and $0\le\beta=v_{t}/v_y<1$, one obtains
\begin{align}
\mathrm{Re}\sigma_{jj(\mathrm{inter})}^{\kappa s}(\omega)
&=\int\frac{d\tilde{k}_x d\tilde{k}_y}{4\pi v_xv_y}\sum_{s^\prime=\pm}
\tilde{\mathcal{F}}_{jj;ss^\prime}^{\kappa;-,+}(\tilde{k}_x,\tilde{k}_y)
\frac{f[\tilde{\varepsilon}_{\kappa s}^{-}(\tilde{k}_x,\tilde{k}_y)]
-f[\tilde{\varepsilon}_{\kappa s^\prime}^{+}(\tilde{k}_x,\tilde{k}_y)]}{\omega}
\delta\left[\omega-2\tilde{k}\right],
\end{align}
where
\begin{align}
\tilde{\mathcal{F}}_{xx;ss^\prime}^{\kappa;-,+}(\tilde{k}_x,\tilde{k}_y) &=
\mathcal{F}_{xx;ss^\prime}^{\kappa;-,+}(k_x,k_y)
=4\sigma_0\frac{v_x^2\tilde{k}_y^2}{\tilde{k}^2}\delta_{ss^\prime},\\
\tilde{\mathcal{F}}_{yy;ss^\prime}^{\kappa;-,+}(\tilde{k}_x,\tilde{k}_y) &=
\mathcal{F}_{yy;ss^\prime}^{\kappa;-,+}(k_x,k_y)
=4\sigma_0\frac{v_y^2\tilde{k}_x^2}{\tilde{k}^2}\delta_{ss^\prime},\\
\tilde{\varepsilon}_{\kappa s}^{\pm}(\tilde{k}_x,\tilde{k}_y)
&=\varepsilon_{\kappa }^{\pm}(k_x,k_y)
=-\kappa\beta\tilde{k}_y\pm\tilde{k}=(-\kappa\beta\sin\phi\pm1)\tilde{k}.
\end{align}

Summing over $s^\prime$ and transforming to the polar coordinate, we have
\begin{align}
\mathrm{Re}\sigma_{xx(\mathrm{inter})}^{\kappa s}(\omega)
&=\sigma_0\frac{v_x}{v_y}\int_{0}^{+\infty}\tilde{k}d\tilde{k}
\int_{0}^{2\pi}\frac{\sin^2\phi d\phi}{\pi}
\frac{f[\tilde{\varepsilon}_{\kappa s}^{-}(\tilde{k}\cos\phi,\tilde{k}\sin\phi)]
-f[\tilde{\varepsilon}_{\kappa s}^{+}(\tilde{k}\cos\phi,\tilde{k}\sin\phi)]}{\omega}
\delta\left[\omega-2\tilde{k}\right],
\end{align}
and
\begin{align}
\mathrm{Re}\sigma_{yy(\mathrm{inter})}^{\kappa s}(\omega)
&=\sigma_0\frac{v_y}{v_x}\int_{0}^{+\infty}\tilde{k}d\tilde{k}
\int_{0}^{2\pi}\frac{\cos^2\phi d\phi}{\pi}
\frac{f[\tilde{\varepsilon}_{\kappa s}^{-}(\tilde{k}\cos\phi,\tilde{k}\sin\phi)]
-f[\tilde{\varepsilon}_{\kappa s}^{+}(\tilde{k}\cos\phi,\tilde{k}\sin\phi)]}{\omega}
\delta\left[\omega-2\tilde{k}\right].
\end{align}

Integrating over $\tilde{k}$ leads us to
\begin{align}
\mathrm{Re}\sigma_{xx(\mathrm{inter})}^{\kappa s}(\omega)
&=\frac{\sigma_0}{4}\frac{v_x}{v_y}
\int_{0}^{2\pi}\frac{\sin^2\phi d\phi}{\pi}
\left\{f\left[(-\kappa\beta \sin\phi-1)\frac{\omega}{2}\right]
-f\left[(-\kappa\beta \sin\phi+1)\frac{\omega}{2}\right]\right\}\nonumber\\
&=\frac{\sigma_0}{4}\frac{v_x}{v_y}
\int_{0}^{\pi}\frac{\sin^2\phi d\phi}{\pi}
\left\{f\left[(-\kappa\beta \sin\phi-1)\frac{\omega}{2}\right]
-f\left[(-\kappa\beta \sin\phi+1)\frac{\omega}{2}\right]
\right.\nonumber\\&\hspace{3.25cm}\left.
+f\left[(\kappa\beta \sin\phi-1)\frac{\omega}{2}\right]
-f\left[(\kappa\beta \sin\phi+1)\frac{\omega}{2}\right]\right\}
\end{align}
and hence
\begin{align}
\mathrm{Re}\sigma_{xx(\mathrm{inter})}(\omega)
&=g_s\left[\mathrm{Re}\sigma_{xx(\mathrm{inter})}^{+}(\omega)+\mathrm{Re}\sigma_{xx(\mathrm{inter})}^{-}(\omega)\right]\notag\\
&=\sigma_0\frac{v_x}{v_y}
\int_{0}^{\pi}\frac{\sin^2\phi d\phi}{\pi}\left\{f\left[(-\beta \sin\phi-1)\frac{\omega}{2}\right]
+f\left[(\beta \sin\phi-1)\frac{\omega}{2}\right]
\right.\nonumber\\&\hspace{3.2cm}\left.
-f\left[(-\beta \sin\phi+1)\frac{\omega}{2}\right]
-f\left[(\beta \sin\phi+1)\frac{\omega}{2}\right]\right\},
\end{align}
where $g_s=2$ is the degeneracy factor of spin.

Parallel procedures give rise to
\begin{align}
\mathrm{Re}\sigma_{yy(\mathrm{inter})}(\omega)
&=g_s\left[\mathrm{Re}\sigma_{yy(\mathrm{inter})}^{+}(\omega)+\mathrm{Re}\sigma_{yy(\mathrm{inter})}^{-}(\omega)\right]\notag\\
&=\sigma_0\frac{v_y}{v_x}
\int_{0}^{\pi}\frac{\cos^2\phi d\phi}{\pi}\left\{f\left[(-\beta \sin\phi-1)\frac{\omega}{2}\right]
+f\left[(\beta \sin\phi-1)\frac{\omega}{2}\right]
\right.\nonumber\\&\hspace{3.2cm}\left.
-f\left[(-\beta \sin\phi+1)\frac{\omega}{2}\right]
-f\left[(\beta \sin\phi+1)\frac{\omega}{2}\right]\right\}.
\end{align}

In order to obtain the analytical expressions, we perform the integrations over $\phi$ at zero temperature where the Fermi distribution function $f(x)$ can be replaced by $\Theta[\mu-x]$. 
At zero temperature, we have
\begin{align}
\mathrm{Re}\sigma_{xx(\mathrm{inter})}(\omega)
&=\sigma_0\frac{v_x}{v_y}\Gamma_\beta(\omega),
\end{align}
where
\begin{align}
\Gamma_\beta(\omega)
&=\Theta\left(\omega\right)\int_{0}^{\pi}\frac{\sin^2\phi d\phi}{\pi}
\left\{\Theta\left[\mu+(\beta \sin\phi+1)\frac{\omega}{2}\right]
+\Theta\left[\mu-(\beta \sin\phi-1)\frac{\omega}{2}\right]
\right.\nonumber\\&\left.\hspace{3.2cm}
-\Theta\left[\mu+(\beta \sin\phi-1)\frac{\omega}{2}\right]
-\Theta\left[\mu-(\beta \sin\phi+1)\frac{\omega}{2}\right]\right\}.
\end{align}

When $\beta=0$, we have
\begin{align}
\Gamma_0(\omega)
&=\Theta\left(\omega\right)\int_{0}^{\pi}\frac{2\sin^2\phi}{\pi} d\phi\left[
\Theta\left(\mu+\frac{\omega}{2}\right)-\Theta\left(\mu-\frac{\omega}{2}\right)\right]
=
\begin{cases}
0, &\hspace{1cm} 0<\omega<2|\mu|;\\
1, &\hspace{1cm} \omega\ge2|\mu|.
\end{cases}
\end{align}

When $0<\beta<1$, we have
\begin{align}
\Gamma_\beta(\omega)
&=\Theta\left(\beta\right)\Theta\left(1-\beta\right)\Theta\left(\omega\right)
\int_{0}^{\frac{\pi}{2}}\frac{2\sin^2\phi \cos\phi d\phi}{\pi\cos\phi}
\left\{\Theta\left[\mu+(\beta \sin\phi+1)\frac{\omega}{2}\right]
+\Theta\left[\mu-(\beta \sin\phi-1)\frac{\omega}{2}\right]
\right.\nonumber\\&\left.\hspace{6.5cm}
-\Theta\left[\mu+(\beta \sin\phi-1)\frac{\omega}{2}\right]
-\Theta\left[\mu-(\beta \sin\phi+1)\frac{\omega}{2}\right]\right\}\notag\\
&=\Theta\left(\beta\right)\Theta\left(1-\beta\right)\Theta\left(\omega\right)
\int_{0}^{1}\frac{2x^2 dx}{\pi\sqrt{1-x^2}}
\left\{\Theta\left[\mu+(\beta x+1)\frac{\omega}{2}\right]
+\Theta\left[\mu-(\beta x-1)\frac{\omega}{2}\right]
\right.\nonumber\\&\left.\hspace{5.8cm}
-\Theta\left[\mu+(\beta x-1)\frac{\omega}{2}\right]
-\Theta\left[\mu-(\beta x+1)\frac{\omega}{2}\right]\right\}\notag\\
&=
\begin{cases}
0, &\hspace{1cm} 0<\omega<\frac{2|\mu|}{1+\beta};\\
\frac{1}{2}-\frac{\chi\sqrt{1-\chi^2}-\arcsin\chi}{\pi}, &\hspace{1cm} \frac{2|\mu|}{1+\beta}\le\omega<\frac{2|\mu|}{1-\beta},\\
1, &\hspace{1cm} \omega\ge\frac{2|\mu|}{1-\beta},
\end{cases}
\end{align}
where $\chi=\frac{\omega-2|\mu|}{\beta\omega}$. Therefore the analytical expressions of $\mathrm{Re}\sigma_{xx(\mathrm{inter})}(\omega)$ at zero temperature can be written in the untilted case ($\beta=0$)
as
\begin{align}
\mathrm{Re}\sigma_{xx(\mathrm{inter})}(\omega)=
\begin{cases}
0, &\hspace{1cm} 0<\omega<2|\mu|;\\
\sigma_0\frac{v_x}{v_y}, &\hspace{1cm} \omega\ge2|\mu|,
\end{cases}
\end{align}
and in the tilted case ($0<\beta<1$) as
\begin{align}
\mathrm{Re}\sigma_{xx(\mathrm{inter})}(\omega)=
\begin{cases}
0, &\hspace{1cm} 0<\omega<\frac{2|\mu|}{1+\beta};\\
\sigma_0\frac{v_x}{v_y} \left[\frac{1}{2}-\frac{\chi\sqrt{1-\chi^2}-\arcsin\chi}{\pi}\right],
&\hspace{1cm} \frac{2|\mu|}{1+\beta}\le\omega<\frac{2|\mu|}{1-\beta},\\
\sigma_0\frac{v_x}{v_y}, &\hspace{1cm} \omega\ge\frac{2|\mu|}{1-\beta},
\end{cases}
\label{Analyticalxx}
\end{align}
where $\chi=\frac{\omega-2|\mu|}{\beta\omega}$.

Similarly, the analytical expressions of $\mathrm{Re}\sigma_{yy(\mathrm{inter})}(\omega)$ at zero temperature can be written in the untilted case ($\beta=0$) as
\begin{align}
\mathrm{Re}\sigma_{yy(\mathrm{inter})}(\omega)
=
\begin{cases}
0, &\hspace{1cm} 0<\omega<2|\mu|;\\
\sigma_0\frac{v_y}{v_x}, &\hspace{1cm} \omega\ge2|\mu|,
\end{cases}
\end{align}
and in the tilted case ($0<\beta<1$) as
\begin{align}
\mathrm{Re}\sigma_{yy(\mathrm{inter})}(\omega)&=
\begin{cases}
0, &\hspace{1cm} 0<\omega<\frac{2|\mu|}{1+\beta},\\
\sigma_0\frac{v_y}{v_x} \left[\frac{1}{2}+\frac{\chi\sqrt{1-\chi^2}+\arcsin\chi}{\pi}\right],
&\hspace{1cm} \frac{2|\mu|}{1+\beta}\le\omega<\frac{2|\mu|}{1-\beta},\\
\sigma_0\frac{v_y}{v_x}, &\hspace{1cm} \omega\ge\frac{2|\mu|}{1-\beta},
\end{cases}
\label{Analyticalyy}
\end{align}
where $ \chi=\frac{\omega-2|\mu|}{\beta\omega}$.

It is noted that two boundaries $\frac{2|\mu|}{1+\beta}$ and $\frac{2|\mu|}{1-\beta}$ correspond to the position
of peaks of optical conductivities and are determined only by the tilt parameter $\beta=v_t/v_y$ and the absolute
value of chemical potential $|\mu|$, indicating the tilt dependence and particle-hole symmetry in the gapless model.
In this sense, we analytically evaluate the real part of longitudinal optical conductivities which agree exactly with the numerical results of 8-$Pmmn$ borophene \cite{PRBVerma2017} after setting $v_{-}=v_{t}$, $v_{2}=0$, $v_{1}=v_{x}$, and $v_{+}=v_{y}$. Furthermore, in the untilted limit $\beta\to0$, these two boundaries merge into one boundary $2|\mu|$, giving rise to the result for the untilted case ($\beta=0$).

When $\omega>\frac{2|\mu|}{1-\beta}$ is satisfied, the real part of longitudinal optical conductivities becomes constant, namely,
\begin{align}
\mathrm{Re}\sigma_{xx(\mathrm{asyp})}&=\frac{v_x}{v_y}\sigma_0=\frac{v_1}{\sqrt{v_2^2+v_{+}^2}}\sigma_0,\label{asypxx}\\
\mathrm{Re}\sigma_{yy(\mathrm{asyp})}&=\frac{v_y}{v_x}\sigma_0=\frac{\sqrt{v_2^2+v_{+}^2}}{v_1}\sigma_0,\label{asypyy}
\end{align}
which are nothing but the asymptotic background values. It is noted that they are related only to the ratio between $v_x$ and $v_y$, irrelevant with the tilt parameter $\beta=v_t/v_y$. In addition, they satisfy a universal relation
\begin{align}
\mathrm{Re}\sigma_{xx(\mathrm{asyp})}\times\mathrm{Re}\sigma_{yy(\mathrm{asyp})}
&=\frac{v_x}{v_y}\sigma_0\times\frac{v_y}{v_x}\sigma_0=\sigma_0^2.
\label{asyprelation}
\end{align}
It is remarked that Eqs.(\ref{asypxx}-\ref{asyprelation}) hold for arbitrary chemical potential such that the $n$-doped ($\mu>0$), $p$-doped ($\mu<0$), and undoped ($\mu=0$) cases share the same asymptotic background values. Furthermore, from the Fermi distribution function $f\left[\varepsilon_{\kappa}^{\xi}(k_x,k_y)\right]
=1/\left\{1+\exp\left[\left(\varepsilon_{\kappa}^{\xi}(k_x,k_y)-\mu\right)/(k_B T)\right]\right\}$, when $\varepsilon_{\kappa}^{\xi}(k_x,k_y)$ is much greater than chemical potential $\mu$ and temperature $k_B T$, we are allowed to set $T=0$, which indicates that finite temperature does not affect the asymptotic background values. The result that asymptotic background values of longitudinal optical conductivity are not affected by finite temperature was also found by Carbotte in the numerical calculation of type-I and type-II Weyl semimetals \cite{PRBCarbotte2016}. In summary, both the asymptotic values in Eqs.(\ref{asypxx}) and (\ref{asypyy}) and the universal relation in Eq.(\ref{asyprelation}) hold for arbitrary band gap, arbitrary chemical potential, and arbitrary temperature
since physically the band gap, chemical potential, and temperature are relatively small compared to the energy in the asymptotic regime.

As a further comparison, we adopt the notations $\epsilon_1=\frac{2\mu}{1+\beta}$ and $\epsilon_2=\frac{2\mu}{1-\beta}$ \cite{PRBVerma2017} and can express the ratio $\beta=v_t/v_y$ as $\mu\left(\frac{1}{\epsilon_1}-\frac{1}{\epsilon_2}\right)$ or $\frac{\epsilon_2-\epsilon_1}{\epsilon_2+\epsilon_1}$. Note that the relation $\beta=\mu\left(\frac{1}{\epsilon_1}-\frac{1}{\epsilon_2}\right)$ is nothing but the result given in Ref. \cite{PRBVerma2017}.

\end{widetext}


\end{document}